\documentstyle[latexsym,multicol,aps,prl]{revtex} 

\renewcommand{\narrowtext}{\begin{multicols}{2} \global\columnwidth20.5pc}
\def\be{\begin{equation}}
\def\ee{\end{equation}}
\def\bea{\begin{eqnarray}}
\def\eea{\end{eqnarray}}
\def\bes{\begin{eqnarray*}}
\def\ees{\end{eqnarray*}}
\def\bb{\bibitem}        
\def\ra{\rightarrow}

\def\non{\nonumber}
\def\lapp{\hbox{$ {     \lower.40ex\hbox{$<$}
                   \atop \raise.20ex\hbox{$\sim$}
                   }     $}  }
\def\rapp{\hbox{$ {     \lower.40ex\hbox{$>$}
                   \atop \raise.20ex\hbox{$\sim$}
                   }     $}  }

\begin{document}

\title{
Sharpening Low-Energy, Standard-Model Tests
\\
via Correlation Coefficients in 
Neutron $\beta$-Decay}

\author{S. Gardner\thanks{E-mail: gardner@pa.uky.edu} 
and C. Zhang}
\address{Department of Physics and Astronomy, \\ University of Kentucky,
        Lexington, KY 40506-0055}

\date{March 21, 2001}

\maketitle

\begin{abstract}
The correlation coefficients $a$, $A$, and $B$ in neutron
$\beta$-decay are proportional to the ratio of the 
axial-vector to vector weak coupling 
constants, $g_A/g_V$, to leading recoil order.
With the advent of the next generation of neutron decay
experiments, the recoil-order corrections to these expressions 
become experimentally accessible, admitting 
a plurality of 
Standard Model (SM) tests. The measurement of both $a$ and $A$,
e.g., allows one to test the 
conserved-vector-current (CVC) hypothesis and to search 
for second-class currents (SCC) independently. 
The anticipated precision
of these measurements suggests that the 
bounds on CVC violation and SCC from studies of nuclear $\beta$-decay 
can be qualitatively bettered. 
Departures from SM expectations can be interpreted as
evidence for non-$V-A$ currents. 
\end{abstract}

\narrowtext 

Precision nuclear $\beta$-decay measurements have played
an important role in the rise of the Standard Model (SM),  
giving strong credence to 
the conserved-vector-current (CVC) hypothesis, as well as to
the absence of second-class currents (SCC). 
We show that upcoming neutron-decay experiments can 
sharpen tests of the CVC hypothesis and of the absence of
SCC significantly, eliminating 
assumptions inherent to the nuclear studies.

Searches for CVC violation and SCC
in nuclear $\beta$-decay experiments
have spanned decades of effort. 
We consider a CVC test originally suggested by 
Gell-Mann~\cite{gell58}: 
the strength of the ``weak magnetism'' term of the nucleon weak 
current ought be given by the strength of the corresponding 
electromagnetic M1 transition. The SM test realized from 
such a comparison constrains a combination of the weak
magnetism and induced tensor terms of the nucleon weak current. 
The induced tensor term is a ``second class'' current 
and thus is zero in the SM~\cite{Weinberg:1958ut},
save for isospin-violating effects engendered by the
differing mass and charge of the $u$ and $d$ quarks. 
In tests of this sort, 
the CVC hypothesis is tested if SCC are assumed to be 
zero, or, alternatively, 
the non-existence
of SCC is tested if the CVC hypothesis is assumed to be valid.

Historically, the best constraints on the non-existence of SCC
and CVC violation are realized in the 
mass 12 system~\cite{Commins:1983ns,Grenacs:1985da}.
The CVC hypothesis can be tested through the comparison of 
the spectral shape correction parameters 
$a_{\mp}$ measured in $^{12}{\rm B}\rightarrow ^{12}\!\!{\rm C}$ and
$^{12}{\rm N}\rightarrow ^{12}\!\!{\rm C}$ transitions with 
the strength of the electromagnetic M1 transition from the analog
state of $^{12}{\rm C}$. 
This procedure yields a test of the CVC hypothesis at the 10\% 
level~\cite{Commins:1983ns,Grenacs:1985da,DeBraeckeleer:1992jb}.
In order to realize a SCC test, 
the decays of spin-aligned $^{12}{\rm B}$ and $^{12}{\rm N}$ nuclei are
studied. For purely aligned $1^+ \rightarrow 0^+$ 
transitions~\cite{remalign}, 
the $e^{\mp}$ angular distribution for 
$^{12}{\rm B}$ ($-$) and $^{12}{\rm N}$ ($+$) decay
is given by~\cite{Grenacs:1985da} 
\bea
W_{\mp} (E_e, \theta, {\cal A})
&\propto& p_e E_e(E_e- E_e^{\rm max})^2 
[1 + {\cal A} \alpha_{\mp} P_2(\cos\theta) ] \nonumber ,
\eea
where $p_e$ and $E_e$ are the momentum and energy of the
electron (positron), $E_e^{\rm max}$ is the endpoint energy, 
$\theta$ is the angle between ${\bf p_e}$
and the spin orientation axis, and ${\cal A}$ is the nuclear
alignment.  
The difference $\alpha_- - \alpha_+$ is sensitive to
the weak magnetism term as well as to the induced tensor
term in the nucleon weak current. Unfortunately, 
it is also sensitive to the difference of the axial
charges ($\Delta y \equiv y_+ - y_-$)
in the mirror transitions $^{12}{\rm B} \rightarrow ^{12}\!\!{\rm C}$
and $^{12}{\rm N} \rightarrow ^{12}\!\!{\rm C}$ --- this potentiality 
has been included in only the most recent set of SCC
tests~\cite{Minamisono:1998ag,Minamisono:2000vz}. 
Were $\Delta y=0$ and the experimental weak magnetism
contribution determined from the
M1 electromagnetic transition strength 
from the analog
state of $^{12}{\rm C}$~\cite{kosh79}, as per 
the  CVC hypothesis, 
Refs.~\cite{Minamisono:1998ag}
and \cite{Minamisono:2000vz} would yield
$2M f_T/f_A = 0.12 \pm 0.05 ({\rm stat}) \pm 0.15 ({\rm syst})$ 
and 
$2M f_T/f_A = 0.04 \pm 0.16 ({\rm stat}) \pm 0.04 ({\rm syst})$, 
respectively. Note that $f_T$ and $f_A$ denote the 
induced-tensor
and axial-vector coupling constants 
of the nucleon --- the impulse approximation has been made
in order to relate the nuclear and nucleon weak constants,
note, e.g.,  Ref.~\cite{masu79}. 
This is consistent with the earlier 
result $2M f_T/f_A = -0.21 \pm 0.63$~\cite{masu79}. 
Using $\Delta y = 0.10 \pm 0.05$~\cite{koshi95}, 
Refs.~\cite{Minamisono:1998ag}
and \cite{Minamisono:2000vz} determine that 
$2M f_T/f_A = 0.22 \pm 0.05({\rm stat}) \pm 0.15({\rm syst}) 
\pm 0.05({\rm theor})$ and 
$2M f_T/f_A = 0.14 \pm 0.16({\rm stat}) \pm 0.04({\rm syst}) \pm 0.05 
({\rm theor})$, yielding the combined constraint
$0.01 \le 2M f_T/f_A \le 0.34$ at 90\% CL~\cite{Minamisono:2000vz}. 
This result suggests that $f_T$ is non-zero, 
with a value considerably in excess of SM 
expectations~\cite{Donoghue:1982uk,Shiomi:1996pw}. 
The inferred SCC contribution emerges from assuming the
CVC hypothesis; alternatively, we can assert that SCC
are identically zero in order to ascertain the quantitative validity
of the CVC hypothesis. The uncertainties in
the SCC determination are roughly 5\% of the value of the weak magnetism
contribution, so that the CVC hypothesis is tested to this level.
Note that an analogous test of SCC is possible in the mass 8 system 
as well. 
Combining the radiative decays of the analog doublet
in $^{8}{\rm Be}$~\cite{debrack95,snover00} with measurements of the
$\beta-\alpha$ correlation in $^{8}{\rm Li}\rightarrow^{8}\!\!{\rm Be}$
and $^{8}{\rm B}\rightarrow^{8}\!\!{\rm Be}$ decays~\cite{tribble} 
yields a second-class, induced tensor nuclear form factor which is 
consistent with zero~\cite{snover00}, albeit with an error 
rather larger than in the mass 12 system. 
The mass 8 CVC/SCC studies ought also to suffer
a theoretical correction from the difference in the  
allowed axial matrix elements
in the mirror $^{8}{\rm Li}\rightarrow ^{8}\!\!{\rm Be}$ 
and $^{8}{\rm B}\rightarrow ^{8}\!\!{\rm Be}$ decays; the induced
tensor form factor of Ref.~\cite{snover00} assumes
this correction is zero. 

We believe that a crisper test of the CVC hypothesis and of 
the non-existence of SCC 
can be realized via the empirical determination
of the correlation coefficients of neutron $\beta$-decay. 
Thus far, 
the especial focus of these experiments
has been the determination of the 
Cabibbo-Kobayashi-Maskawa (CKM) matrix element 
$V_{ud}$. The latter is extracted from $g_V$, which is determined
from the neutron-spin--electron-momentum correlation $A$ and the
neutron lifetime $\tau_n$. 
The various determinations of $A$ do
not agree~\cite{Aexp}; 
a scale factor of 1.9 is assigned to 
the determination of $g_A/g_V$ from the measured values of $A$
by Ref.~\cite{Groom:2000in}. 
These measurements were realized in 
reactor beam experiments; $A$ can also be measured
using ultra-cold neutron sources --- the systematic errors
in such experiments 
are very different and would seem to be much smaller~\cite{lanlA}. 
Nevertheless, the extracted value of $V_{ud}$,
in concert
with $V_{us}$ from $K_{l3}$ decays, tests the ``squashed''
unitarity relation $|V_{ud}|^2 + |V_{us}|^2 + |V_{ub}|^2=1$
to better than 1\%.
$V_{ud}$ may also be determined, indeed, more precisely, 
from the ``superallowed''
$0^+ \rightarrow 0^+$ decays in nuclei. In this case the
empirical unitarity relation deviates from unity 
by $2.2\sigma$; it is
worth noting, however, that in this case the 
estimated theoretical errors dominate
the presumed error bar~\cite{Towner:1998qj}.

Let us consider the correlation coefficients in neutron
$\beta$-decay.
The differential decay rate of a free neutron 
is given by~\cite{treiman57}:
\bea
&d\Gamma \propto 
E_e|{\bf p_e}|(E_e^{\rm max}-E_e)^2[1 
    +a\frac{{\bf p_e}\cdot{\bf p_\nu}}{E_eE_\nu} 
    +A\frac{\mathbf{{\cal P}}\cdot{\mathbf{p}}_e}{E_e} \non \\
&+B\frac{\mathbf{{\cal P}}\cdot{\mathbf{p}}_\nu}{E_\nu}
    +D
\frac{{\mathbf{{\cal P}}\cdot(\mathbf{p}_e\times{\mathbf{p}_\nu})}}{E_eE_\nu}]
dE_e d\Omega_e d\Omega_\nu , \label{aABDdef}
\eea
where $\mathbf{{\cal P}}$ denotes the neutron's 
polarization vector.  
The pseudo-T-odd coefficient $D$ is small~\cite{Lising:2000pa}
and can be neglected. Defining
$\lambda\equiv|g_A|/|g_V|$ and neglecting terms of recoil order  
we have in the SM 
\bea
a = \frac{1-\lambda^2}{1+3\lambda^2} \quad &;& \quad 
A = 2\frac{\lambda(1- \lambda)}{1+3\lambda^2} \;, \label{A0def} \\
B &=& 2\frac{\lambda(1+ \lambda)}{1+ 3\lambda^2} \nonumber \;. 
\eea
These relations imply that~\cite{most76}
\bea
&1+A-B-a=0 \;, \label{F1} \\
&aB-A-A^2=0. 
\label{F2}
\eea 
Currently~\cite{Groom:2000in}
\bea
a = -0.102 \pm 0.005 \quad &;& \quad A = -0.1162 \pm 0.0013,  \\
B &=& 0.983 \pm 0.004 \;, \nonumber
\eea
so that Eqs. (\ref{F1}) and (\ref{F2}) are satisfied at the
current level of precision. However, these relations
do not hold once terms of recoil order are included. 
The recoil-order terms are controlled by the dimensionless ratio of 
the electron energy to the neutron rest mass and thus are of
${\cal O}(10^{-3})$, so that 
they impact $a$ and $A$ at the 1\% level. The correlation
coefficient $B$ is much larger, so that the recoil order terms
only become important at the 0.1\% level. 
Consequently we will focus on what can be learned from $a$ and $A$. 
Recent experimental proposals suggest that $A$ and possibly $a$
can be measured to 0.2\% or better~\cite{lanlA,nista}. 
We wish to point out that additional Standard Model tests are possible
once terms of recoil order become empirically accessible. 
In particular, one is sensitive to both the weak magnetism term $f_2$ as well
as to the induced tensor
term $g_2$ in the nucleon 
weak current. Indeed, independent
tests of the CVC hypothesis and of the non-existence of SCC 
are possible, as we shall now see. 

The matrix element for polarized neutron $\beta$-decay in the
SM is given by
  \be
  {\cal   M}=\frac{G_F}{\sqrt{2}}\langle p|J^\mu(0)|\vec{n} \rangle
       \times[\bar{u}_e(p_e)\gamma_\mu(1+\gamma_5)u_\nu(p_\nu)]\;.
  \ee
We adopt the historic ($1+\gamma_5$) sign convention in order
to retain manifest consistency with earlier 
work~\cite{harr60,bile60,bender}. 
Lorentz invariance and translation invariance implies that
the nucleon weak current $\langle p|J^\mu(0)|\vec{n} \rangle$
has six terms:
\bea
\langle p(p')|&&J^\mu(0)|\vec{n}(p,{\bf s})\rangle 
= 
\bar{u}_p(p')[f_1(q^2)\gamma^\mu
-i\frac{f_2(q^2)}{M}\sigma^{\mu\nu}q_\nu \nonumber \\
&& +\frac{f_3(q^2)}{M}q^\mu 
 +g_1(q^2)\gamma^\mu\gamma_5-i\frac{g_2(q^2)}{M}{\sigma^{\mu\nu}}\gamma_5
q_\nu \non \\
&& +\frac{g_3(q^2)}{M}\gamma_5q^\mu]u_n(p,{\bf s}), \label{hadcurr}
\eea
%
where $\sigma^{\mu\nu}=\frac{i}{2}[\gamma^\mu,\gamma^\nu]$
and $q=p-p'$. Note that $f_1(0)=g_V$, $g_1(0)=-g_A=-f_A/G_F$, 
and $g_2(0)=-f_T\,M/G_F$, whereas
$M$ and $M'$ are the neutron and
proton mass, respectively. 
The differential decay rate is given by 
\bea
&d^3\Gamma  = 
\frac{|G_F|^2
}{2(2\pi)^5}
\frac{|\mathbf{p}_e||\mathbf{p}_\nu|}
{M-E_e+|\mathbf{p}_e|\cos\theta}
[C_1+C_2({\mathbf{{\cal P}}}\cdot{\mathbf{p}}_e) \nonumber \\
& +C_3({\mathbf{{\cal P}}}\cdot{\mathbf{p}}_\nu)
+C_4{\mathbf{{\cal P}}\cdot(\mathbf{p}_e\times{\mathbf{p}_\nu})}]
dE_e d\Omega_e d\Omega_\nu \;,
\label{cdef}
\eea                                  
where the coefficients $C_i$ contain the form factors of 
Eq.~(\ref{hadcurr})
and are detailed in  Ref.~\cite{harr60}. 
Note that
$\theta$ is the angle between the electron and neutrino momenta
in the neutron rest frame. 
Our particular interest are the recoil corrections to
$a$ and $A$. Let us first consider the case in which 
the neutron is unpolarized. We have 
\bea
&d^2\Gamma  = 
\frac{2|G_F|^2
|g_V|^2
}{(2\pi)^4}
\frac
{(MR)^4 \beta x^2 (1-x)^2}
{(1-Rx+Rx\beta\cos\theta)^3}
[C_a \non \\ 
&+C_b\beta\cos\theta] dE_ed\Omega_{e\nu}\;,
\eea             
where
\bea
R =\frac{E_e^{\rm max}}{M} =\frac{1}{2}(1 + \epsilon - \eta^2)
\quad &; \quad
    x =\frac{E_e}{E_e^{\rm max}} \;,\\
\eta =\frac{M'}{M} \quad &; \quad \epsilon =(\frac{m_e}{M})^2 
\nonumber
\eea
and $C_a + C_b \cos \theta = C_1/(2ME_\nu E_l |g_V|^2)$ --- $C_1$ 
contains the electron-anti-neutrino correlation, $a$. 
Working in leading recoil order, including the phase space
contributions, we have
\bea
&d^2 \Gamma  = 
\frac{2|G_F|^2|g_V|^2
}{(2\pi)^4}
{(MR)^4 \beta x^2 (1-x)^2}
 [\tilde C_a \non   \\
&+ \tilde C_b\beta\cos\theta 
+ \tilde C_c\beta^2\cos^2\theta] 
dE_ed\Omega_{e\nu}\;, 
\label{ctildef}
\eea
where 
\bea
\tilde C_a &=& 1 + 3 \lambda^2 
-\frac{\epsilon}{Rx} 
(1 + 2\lambda + \lambda^2 + 4\tilde f_2 \lambda + 2 \tilde g_2 \lambda
\nonumber \\
&&- 2 \tilde f_3)  
 - R (2\lambda^2 + 2 \lambda + 4\tilde f_2 \lambda + 4 \lambda \tilde g_2)
\nonumber \\
&& + Rx (3 + 9 \lambda^2 + 4\lambda + 8\tilde f_2 \lambda) \;, \\
\tilde C_b &=& 
1 - \lambda^2 + 
R(2\lambda + 2\lambda^2 + 4\tilde f_2\lambda 
+ 4\lambda \tilde g_2) \nonumber \\
&&-4 Rx(\lambda + 3\lambda^2 + 2\tilde f_2 \lambda) \nonumber \\
\tilde C_c &=& -3Rx(1 - \lambda^2) \; \nonumber
\eea
with $\tilde f_2\equiv f_2(0)/f_1(0)$ and 
$\tilde g_2\equiv g_2(0)/f_1(0)$.
The momentum dependence of the form factors does not appear, as
this effect first enters in
next-to-leading recoil order. 
Noting Eq.~(\ref{aABDdef}) we have
$a = 
\tilde C_b/(\tilde C_a + \tilde C_c \beta^2 \cos^2 \theta)$
and thus 
\bea
a &&= \frac{1-\lambda^2}{1+3\lambda^2}
+\frac{1}{(1+3\lambda^2)^2}\Bigg\{ 
\frac{\epsilon}{Rx} \Big[(1 - \lambda^2)(1 + 2\lambda 
 + \lambda^2 \non \\
&& + 2\lambda \tilde g_2 + 4\lambda \tilde f_2 - 2 \tilde f_3)
\Big] 
 + 4R\Big[ (1+\lambda^2)(\lambda^2 + \lambda  \non \\ 
&& + 2\lambda(\tilde f_2 + \tilde g_2))\Big]      
 - Rx\Big[
3(1+3\lambda^2)^2 + 8\lambda(1+\lambda^2) \non \\ && 
\times(1+2\tilde f_2) + 3(\lambda^2 -1)^2 \beta^2 \cos^2 \theta 
\Big] \Bigg\} + {\cal O}(R^2,\epsilon)\;.
\label{arecoil}
\eea
If $\tilde f_3= \tilde g_2 =0$, this expression 
becomes that of Ref.~\cite{bile60}. Note, too, that 
it is also in agreement with Ref.~\cite{Holstein:1974zf}.

The recoil correction to $A$ is determined from Eq.(\ref{cdef})
by integrating over the neutrino variables. We find~\cite{harr60}
\bea
d^2\Gamma &=& 
\frac{2|G_F|^2 |g_V|^2}{(2\pi)^3}
\frac{\beta x^2(1-x)^2}
{(1 + \epsilon -2Rx)^3}
[C_a'\non \\
&& +C_b'\beta {\cal P}\cos\theta_{\cal P}] 
dE_ed(\cos\theta_{\cal P}) \;,
\eea             
where $\theta_{\cal P}$ is the angle between 
the neutron's polarization vector and the electron 
momentum 
in the neutron 
rest frame. $C_2$ and $C_3$ give rise to $C_b'$, whereas
$C_1$ gives rise to $C_a'$. Noting $A=C_b'/C_a'$, we have 
\bea
A &&= \frac{2\lambda(1 - \lambda)}{1+3\lambda^2}
+\frac{1}{(1+3\lambda^2)^2}\Bigg\{ 
\frac{\epsilon}{Rx} \Big[4\lambda^2(1-\lambda)(1 + \lambda \non \\
& & + 2\tilde f_2)
+ 4 \lambda(1 - \lambda)(\lambda \tilde g_2 - \tilde f_3)
\Big] 
 + R\Big[ 
\frac{2}{3} 
(1+\lambda \nonumber \\ && 
+ 
2(\tilde f_2 + \tilde g_2))(3\lambda^2 + 2\lambda -1) 
\Big]     
  + Rx\Big[
\frac{2}{3} (1+\lambda + 2\tilde f_2) \non \\
&& \times(1 - 5\lambda - 9 \lambda^2  -3\lambda^3) + 
\frac{4}{3} \tilde g_2 (1+\lambda + 3 \lambda^2 + 3 \lambda^3) 
\Big] \Bigg\} \non \\ && + {\cal O}(R^2,\epsilon)\;.
\label{Arecoil}
\eea
If $\tilde f_3= \tilde g_2 =0$, this expression 
becomes that of Refs.~\cite{bile60,Wilkinson:1982hu}.
Our result is also
in agreement with Ref.~\cite{Holstein:1974zf}.
Our results are germane to hyperon decay as well;   
in this context either approximate expressions or 
the $E_e$-integrated asymmetry parameters
are reported~\cite{hyper}. 
The recoil corrections to the correlation coefficients
take the form: $a_0 R  + a_1 Rx + a_{-1} {\epsilon}/{Rx}$.
The energy dependence of the three terms is 
distinct, although only two terms are empirically
accessible as 
$\epsilon/R \sim 2.2 \cdot 10^{-4}$, whereas
$R \sim 1.4 \cdot 10^{-3}$. 
Note that $x \in [\sqrt{\epsilon}/R, 1 ]$. 
Thus we have four independent empirical constraints, i.e., 
the $x^0$ and $x^1$ terms in $a$ and $A$, and three 
unknowns --- namely, 
$\lambda$, $\tilde f_2$, and $\tilde g_2$. The system
is overconstrained, so that we can infer the existence
of physics beyond the SM, namely the presence of 
non-$V-A$ currents~\cite{treiman57}, if the extracted coupling
constants differ from SM bounds or if the values of the extracted
couplings are 
not consistent with each other. Note that independent linear combinations 
of $\tilde f_2$ and
$\tilde g_2$ appear in 
$a$ and $A$, so that, unlike the nuclear cases commonly studied, 
each coupling constant can be determined independently. 
Evaluating the recoil-order contributions to $a$ and
$A$, using $\lambda =1.2670$~{\cite{Groom:2000in}}, 
$\tilde{g_2}=0$, and 
$\tilde{f_2}=(\kappa_p - \kappa_n)/2=1.8529$, as
per the CVC hypothesis, we find that the  
recoil corrections to $a$ are roughly a factor of two larger than
those to $A$. By virtue of the allowed terms, 
$\lambda$ is determined to 0.030\% and 0.022\% by 
0.1\% measurements of $a$ and $A$, respectively. On statistical
grounds, a precision measurement of $A$ would be the most efficacious
in determining $\lambda$, whereas the determination of the 
coupling constants appearing in recoil order would seem to 
be better served with an $a$ measurement. 
$\tilde f_2$ and $\tilde g_2$ can be determined in a 
plurality of ways; let us illustrate. 
Firstly, 
the $x^1$ and $x^0$ terms in $a$ can be determined to yield 
$\tilde f_2$ and $\tilde f_2 + \tilde g_2$. 
$\lambda$ will
be sufficiently precisely determined to have little impact
on the errors in these parameters. Ignoring this source
of error, we find a 0.1\% measurement 
of $a$ yields a 2.5\% measurement of $\tilde f_2$
from the $x^1$ term. 
This, in concert with the $x^0$ term from a 0.1\% measurement of 
$a$, yields an uncertainty in $\tilde g_2$ of
order $0.22\lambda/2$ --- this is compatible with the
errors quoted in the mass 12 experiment with far fewer
assumptions. 
Secondly, the $x^1$ dependence of the $a$ and
$A$ terms can be determined -- the former yields $\tilde f_2$,
whereas the latter yields a combination of $\tilde f_2$  
and $\tilde g_2$~\cite{hols}. 
Earlier determinations of $a$ were inferred from
the recoil proton's spectral shape, see, e.g., Ref.~\cite{Stratowa:1978gq}, 
and were insensitive to the  
$x$ dependence of $a$; the newly proposed
$a$ experiment~\cite{nista} would be the first to measure $a$ as a
function of $x$~\cite{fred}. 
The Fierz interference term, $b$~\cite{treiman57}, 
which is zero in the SM can thus be bounded as well.
Combining the earlier determination of $\tilde f_2$  
with a 0.1\% measurement of 
$A$ to determine $\tilde g_2$ from the $x^1$ term 
yields an uncertainty of
$0.26\lambda/2$, commensurate with our earlier estimate.
Although 0.1\% measurements of $A$ seem 
quite feasible~\cite{brad}, measurements of $a$ to better
than 1\% may pose an especial challenge. Nevertheless, 
precision measurements of $a$ and $A$ are richly complementary.
The measurement of both $a$ and $A$ permit crisp 
SM tests, namely of SCC and the CVC hypothesis, 
not realizable in nuclear decays.

We thank M.~S. Dewey and F. Wietfeldt for introducing us to
the NIST ``$a$'' experiment, and 
we gratefully acknowledge F. Wietfeldt, E. Adelberger, 
B. Filippone, B. Holstein, R. McKeown, K. Snover, and
J. Tandean 
for additional helpful conversations and correspondence. 
We acknowledge the support of the DOE under contract 
number DE-FG02-96ER40989.

\end{multicols} 

\end{document}